\documentclass[runningheads]{svmult}

\usepackage{makeidx}   
\usepackage{graphicx}  
\usepackage{subeqnar}  
\usepackage{multicol}  
\usepackage{physprbb}  
\makeindex             

\begin{document}
\title*{The Luminosity Function of the Host Galaxies of\protect\newline QSOs and BL Lac Objects}
\toctitle{The Luminosity Function of the Host Galaxies of\protect\newline QSOs and BL Lac Objects}
%
%
\titlerunning{The HGLF of QSOs and BL Lacs}
%
\author{Nicoletta Carangelo\inst{1}
\and Renato Falomo\inst{2}
\and Aldo Treves\inst{1}}
\authorrunning{Nicoletta Carangelo et al.}
%
%
\institute{Universit\`a dell'Insubria, via Valleggio 11, Como, Italy
\and Osservatorio Astronomico di Padova,
     Vicolo dell'Osservatorio 5,
     Padova,Italy}

\maketitle              

\section {Introduction}
A clear insight of the galaxies hosting active galactic nuclei is of fundamental importance for understanding the processes of galaxies and nuclei formation and their cosmic evolution. A good characterization of the host galaxies properties requires images of excellent quality in order to disentangle the light of the galaxy from that of the bright nucleus. To this aim HST has provided a major improvement of data on QSOs (Disney et al. 1995; Bahcall et al. 1996, 1997; Boyce et al. 1998; McLure et al. 1999; Hamilton et al. 2000; 
Kukula et al. 2001) and BL Lacs (Scarpa et al. 2000, Urry et al. 2000).\\

We present a comparative study of low redshift QSO and BL Lac host galaxy luminosity function (HGLF). To this aim we have considered samples of BL Lacs (Urry et al. 2000) and QSOs (Bahcall et al. 1997; Boyce et al. 1998; McLure et al. 1999) that have been well resolved by images obtained with WFPC2 on board of HST.
 
\section{The datasets}
We have collected data for BL Lacs and QSOs at z$<$0.5 observed with WFPC2 of HST.
All magnitudes are converted into R (Cousins) band;
absolute magnitudes are calculated assuming for H$_{0}$=50 Km s$^{-1}$ Mpc$^{-1}$ and $\Omega_{0}$=0 and applying uniform k and galactic reddening corrections.\\

The HST snapshot imaging survey of BL Lacs (Urry et al. 2000, Scarpa et al. 2000) has provided a homogeneous set of 110 short exposure high resolution images through the F702W filter. From this dataset we have considered all objects at z$<$0.5 that are resolved in the HST images. This yields 57 sources with 0.027$<z<$0.495 ($<z>$=0.2$\pm$0.1). For these objects the associated host galaxy morphology is always well described by de Vaucouleurs modelling.\\

There is not a comparable large set of HST observation for QSOs, therefore we have considered two representative datasets (Hamilton et al. 2000 and Treves et al. 2001) constructed from a collection of various sources reporting QSO images secured by HST.
Hamilton et al. 2000 have investigated HST archival images of 71 QSOs (26 RLQs and 45 RQQs) with M$_{V}\leq$-23 mag and redshift 0.06$\leq z \leq$0.46.
Treves et al. 2001 have reported on 15 RLQs (in the redshift in the range 0.158$<z<$0.389) collected from the samples of Bahcall et al. 1997, McLure et al. 1999 and Boyce et al. 1998, and homogeneized to the sample of BL Lacs (see Treves et al. 2001). The average luminosities of the above samples are summarized in Table 1.\\

\begin{table}
\caption{Properties of the datasets}
\begin{center}
\renewcommand{\arraystretch}{0.5}
\setlength\tabcolsep{5pt}
\begin{tabular}{llcccc}
\hline\noalign{\smallskip}
Dataset & Reference & N$_{obj}$ & $<z>$ & $<M_{B}(nuc)>$ & $<M_{R}(host)>$\\
\hline
\noalign{\smallskip}
 BL Lacs & Urry et al. 2000 & 57 & 0.20 & -22.3 & -23.7  \\
 RLQs & Treves et al. 2001 &  15 & 0.26 & -25.1 & -24.3  \\
 RQQs & Hamilton et al. 2000 & 45 & 0.22 & -23.9 & -23.8\\ 
 RLQs & Hamilton et al. 2000 & 26 & 0.29 & -25.3 & -24.8\\
\hline
\end{tabular}
\end{center}
\label{Tab1a}
\end{table}

\section {The Host Galaxy Luminosity Function (HGLF)}

Assuming the host galaxy luminosity is independent of nuclear luminosity we consider that the present datasets are representative of the general population of host galaxies of the respective classes and therefore apt to produce a rough luminosity function of the host galaxies.
To set the normalization factor of HGLF for QSOs we took the value of the QSO luminosity function (Boyle et al. 2000) corresponding to the average value of nuclear magnitude in B band (taking B-R$\sim$0.56) and  assumed that RLQs are 10\% of QSO population (Kellermann et al. 1989) at M$_{B}$(nuc)=-25.3.
For BL Lacs we refer to the FRI luminosity function given by Padovani et al. 1991 and normalized the HGLF at M$_{R}$(host)=-22.8. We fit the luminosity function of the host galaxies with a modified Schechter function $\Phi$=K $\times$ $\Phi_{S}$ $\times$ (L/L$^{*}$)$^{\beta}$, where $\Phi_{S}$ is the Schechter function for elliptical galaxies (Metcalfe et al. 1998): $\Phi_{S}$=$\Phi^{*}$ $\times$ (L/L$^{*}$)$^{\alpha}$ $\times$ exp(-L/L$^{*}$), with $\Phi^{*}=8.5\times10^{-2}$ Mpc$^{-3}$, $\alpha$=-1.2 and L$^{*}$=2.25 $\times$ 10$^{44}$ erg s$^{-1}$. The best fit has been estimated minimizing $\chi^{2}$ for the function $\Phi$. We find $\beta\sim$3 for BL Lac and $\beta\sim$5 for RLQ hosts.
The derived HGLFs are given in Fig. 1.\\

\begin{figure}[]
\begin{center}
\includegraphics[width=.5\textwidth]{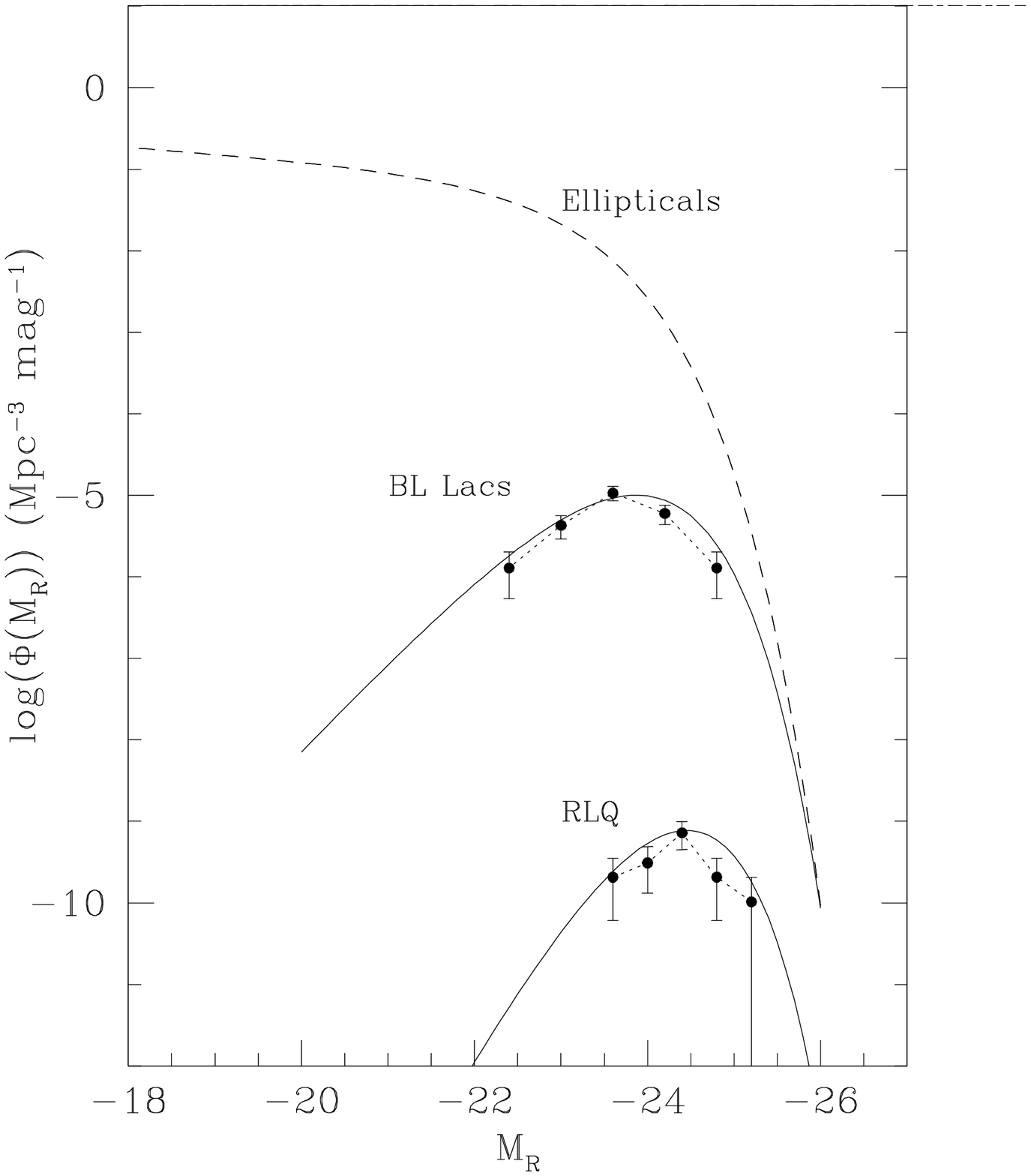}\includegraphics[width=.5\textwidth]{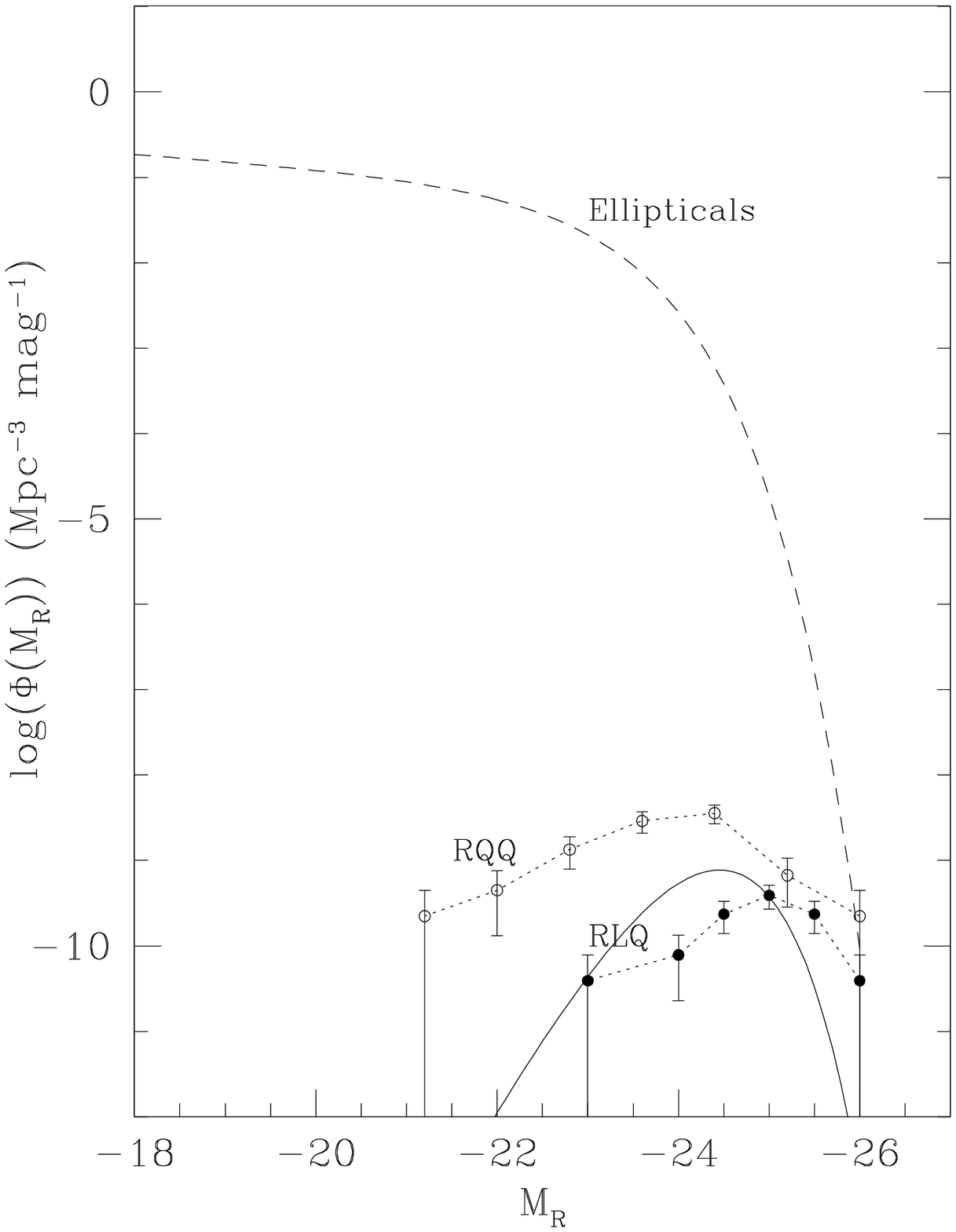}
\end{center}
\caption[]{(\textbf{Left}) The HGLF of RLQs (Treves et al. 2001) and BL Lacs (Urry et al. 2000). The {\it dashed curve} is the elliptical galaxy luminosity function of Metcalfe et al. 1998 and the {\it solid lines} are the fits of the HGLF of RLQs and BL Lacs. (\textbf{Right})
The HGLF of RQQs (empty dot) and RLQs (filled dot) of Hamilton et al. 2000 compared to the luminosity function of Metcalfe et al. ({\it dashed curve}). 
The {\it solid line} is the fit with a modified Schechter function $\Phi$ for the HGLF of the Treves et al. (2001) sample }
\label{eps1}
\end{figure}

\section{Main Conclusions}
\begin {itemize}
\item The HGLFs of QSOs and BL Lacs are remarkably different in shape from the one of inactive ellipticals and indicates that these AGNs are preferentially drawn from the bright tail of elliptical galaxy luminosity function.
\item The HGLFs of RLQs and BL Lacs have similar shape but with a trend for brighter galaxies in RLQs. This is quantified by comparison of the $\beta$ parameter: $\beta_{RLQ}\sim5>\beta_{BLL}\sim3$.
\item There is some indication of a different shape for the HGLFs of RLQs and RQQs with a larger fraction of low luminosity galaxies in RQQs.
\end {itemize}

%

\end{document}